\documentclass[aps,pre,twocolumn,showpacs,preprintnumbers,superscriptaddress]{revtex4}
\usepackage{graphicx}
\usepackage{epsfig}
\usepackage{bm}

\begin{document} 
\title{Sticky grains do not change the universality class of isotropic sandpiles}

\author{Juan A. Bonachela}
\affiliation{Instituto~de~F\'\i sica~Te\'orica~y~Computacional~Carlos~I,
~Facultad~de~Ciencias,~Universidad~de~Granada,~18071~Granada,~Spain}
\author{Jos\'e J. Ramasco}
\affiliation{Physics Department, Emory  University, Atlanta, Georgia 30322, 
USA}
\author{Hugues Chat\'e}
\affiliation{CEA -- Service de Physique de l'\'Etat Condens\'e,~CEN
~Saclay,~91191~Gif-sur-Yvette,~France}
\author{Ivan Dornic}
\affiliation{CEA -- Service de Physique de l'\'Etat Condens\'e,~CEN
~Saclay,~91191~Gif-sur-Yvette,~France}
\affiliation{Instituto~de~F\'\i sica~Te\'orica~y~Computacional~Carlos~I,
~Facultad~de~Ciencias,~Universidad~de~Granada,~18071~Granada,~Spain}
\author{Miguel A. Mu\~noz}
\affiliation{Instituto~de~F\'\i sica~Te\'orica~y~Computacional~Carlos~I,
~Facultad~de~Ciencias,~Universidad~de~Granada,~18071~Granada,~Spain}
\date{\today}

\begin{abstract}
We revisit the sandpile model with ``sticky'' grains introduced by
Mohanty and Dhar [Phys. Rev. Lett. {\bf 89}, 104303 (2002)] whose
scaling properties were claimed to be in the universality class of
directed percolation for both isotropic and directed models.
Simulations in the so-called fixed-energy ensemble show that this
conclusion is not valid for isotropic sandpiles and that this model
shares the same critical properties of other stochastic sandpiles,
such as the Manna model.
These results are strengthened by the analysis of the Langevin
equations proposed by the same authors to account for this problem
which we show to converge, upon coarse-graining, to the
well-established set of Langevin equations for the Manna class.
Therefore, the presence of a conservation law keeps isotropic
sandpiles, with or without stickiness, away from the directed
percolation class.
\end{abstract}

\maketitle

\section{Introduction}

Toy models of sandpiles are the archetypical examples of
self-organized criticality emerging out of time-scale separation
\cite{BTW,Jensen,Dhar}. Sandpile models come in many different flavors
(deterministic or stochastic rules \cite{Manna}, discrete or
continuous variables \cite{Zhang}, with or without height restrictions
\cite{restricted}, etc.), but they usually consist in adding grains one
by one until a local threshold (typically a condition on some slope or
height) is reached, triggering a series of redistribution events, i.e.
``avalanches'', which may lead to dissipation of sandgrains at the
open boundaries.  Their numerical study is notoriously difficult and
first led to a largely unsatisfactory situation in which ``microscopic
details'' were believed to influence scaling properties, in
contradiction with universality principles
\cite{Pacz,Universality}. Major progress in favor of universality
came when sandpile criticality was put into the broader context of
standard non-equilibrium absorbing-state phase transitions
\cite{Marro,Review_Haye,Review_Granada,Review_Odor}
\cite{FES,Granada,BJP,Mikko,Lubeck,KC}.

Indeed, switching off both dissipation (open boundaries) and driving
(slow addition of grains), the total amount of sand or ``energy'' is
conserved, and becomes a control parameter for these ``fixed energy
sandpiles''. For large amounts of sand, the system is in an active
phase with never-ending relaxation events, while for small energies it
gets trapped with certainty into some absorbing state where all
dynamics ceases (all sites being below threshold). Separating these
two regimes there is a critical energy which was shown
\cite{FES,Granada,BJP} to coincide with the stationary energy-density
in the corresponding original sandpile (with slow-driving and boundary
dissipation). In this way, the exponents characterizing sandpiles can
be related to standard critical exponents in an absorbing-state phase
transition \cite{avalanches} (an alternative route not discussed here
is to map sandpiles into standard pinning-depinning interfacial phase
transitions).

Using this approach, it was determined that stochastic sandpiles
\cite{noteBTW} generally do {\it not} belong to the directed
percolation (DP) class, prominent among absorbing phase transitions,
but to the so-called ``conserved-DP'' or Manna class (hereafter
C-DP/Manna) characterized by the coupling of activity to a static
conserved field directly representing the conservation of sandgrains
\cite{FES,Granada,BJP,Lubeck,Romu1,Romu2}. The field theory or 
mesoscopic Langevin equations describing this class reads:
\begin{equation}
\begin{array}{lll}
\partial_t \rho &=& a\rho-b\rho^{2}+\omega\rho E
+D\nabla^2\rho+\sigma\sqrt{\rho}\eta\\
\partial_{t} E &=& D_{E}\nabla^2\rho
\end{array}
\label{sde_Manna}
\end{equation}
where $\rho({\bf x},t)$ is the activity field (characterizing the
density of grains above threshold), $E({\bf x},t)$ is the
locally-conserved energy field, $a, b, \omega, D, \sigma$, and $D_{E}$
are parameters and $\eta ({\bf x},t)$ is a Gaussian white noise.
Equation (\ref{sde_Manna}) represents a robust and well-established
universality class including not only stochastic sandpiles, but also
some reaction-diffusion systems \cite{Romu1,Romu2,Lubeck}

In a recent Letter however, Mohanty and Dhar \cite{MD} have argued
that generic sandpile models with ``sticky grains'' or ``inertia'',
where some grains remain stable even if the local threshold is passed,
should be in the DP class.  The authors presented convincing numerical
and analytical evidence that indeed this is the case for a {\it
directed} two-dimensional system, which happens to be mappable into an
effective one-dimensional directed site-percolation dynamics. Also,
for isotropic ({\it undirected}) models with stickiness it was claimed
that DP scaling holds.  For this second case the authors presented a
Monte Carlo simulation and, additionally justified their findings by
arguing that the ``right'' set of Langevin equations for inertial
sandpiles should include a coupling of the form $\omega \rho
\Theta(E-\rho-E_c)$ where $\Theta$ is the Heaviside step function 
and $E_c$ is the instability threshold, substituting the bilinear
coupling $\omega \rho E$ in Eq. (\ref{sde_Manna}).

The logic behind such a term is, in principle, reasonable \cite{why}
and it is argued in \cite{MD} that considering this coupling, one
should leave the Manna universality class and return to DP, i.e. the
conservation law should be irrelevant in the presence of
``stickiness'' (see the schematic diagram in figure $4$ of \cite{MD}).

In this paper, we argue that this claim in \cite{MD} is unfounded:
even in the presence of inertia/stickiness, the generic universality
class of (undirected) stochastic sandpiles models remains the
C-DP/Manna one. The paper is organized as follows: we first report on
extensive simulations of the model studied in \cite{MD} in the
fixed-energy ensemble, from which we conclude that isotropic sticky
sandpiles are in the C-DP/Manna class. In section III, we integrate
the set of Langevin equations with a $\Theta$ function coupling, which
we find to be also in the C-DP/Manna class, and we perform a numerical
renormalization treatment to show that these Langevin equations evolve
upon coarse-graining towards Eqs.(\ref{sde_Manna}).

\section{Microscopic ``sticky'' sandpile models}

The model proposed in \cite{MD} is a variation of the Manna model: a
discrete sandpile, defined on a one-dimensional lattice, with a height
threshold $h_{\rm c}$, slow sand addition and dissipation at the
boundaries, but including a (sticking) probability $1-p$ for grains to
remain stable even if they are above threshold. Here we consider only
the limit which possesses a critical point, i.e. the bulk-dissipation
rate is set to zero.  Following the strategy in
\cite{FES,Granada,BJP}, we analyze the ``fixed energy
version'' of the model: we suppress grain addition and boundary
dissipation, fix the total energy to $E$, and use $p$ as a control
parameter. Note that, owing to the existence of a non-vanishing
sticking probability, arbitrarily large heights are allowed. Active
sites (at which $h>h_{\rm c}$) are updated in parallel with the
toppling occurring with probability $p$.

\begin{figure}[t]
\begin{center}
\includegraphics[width=4.2cm]{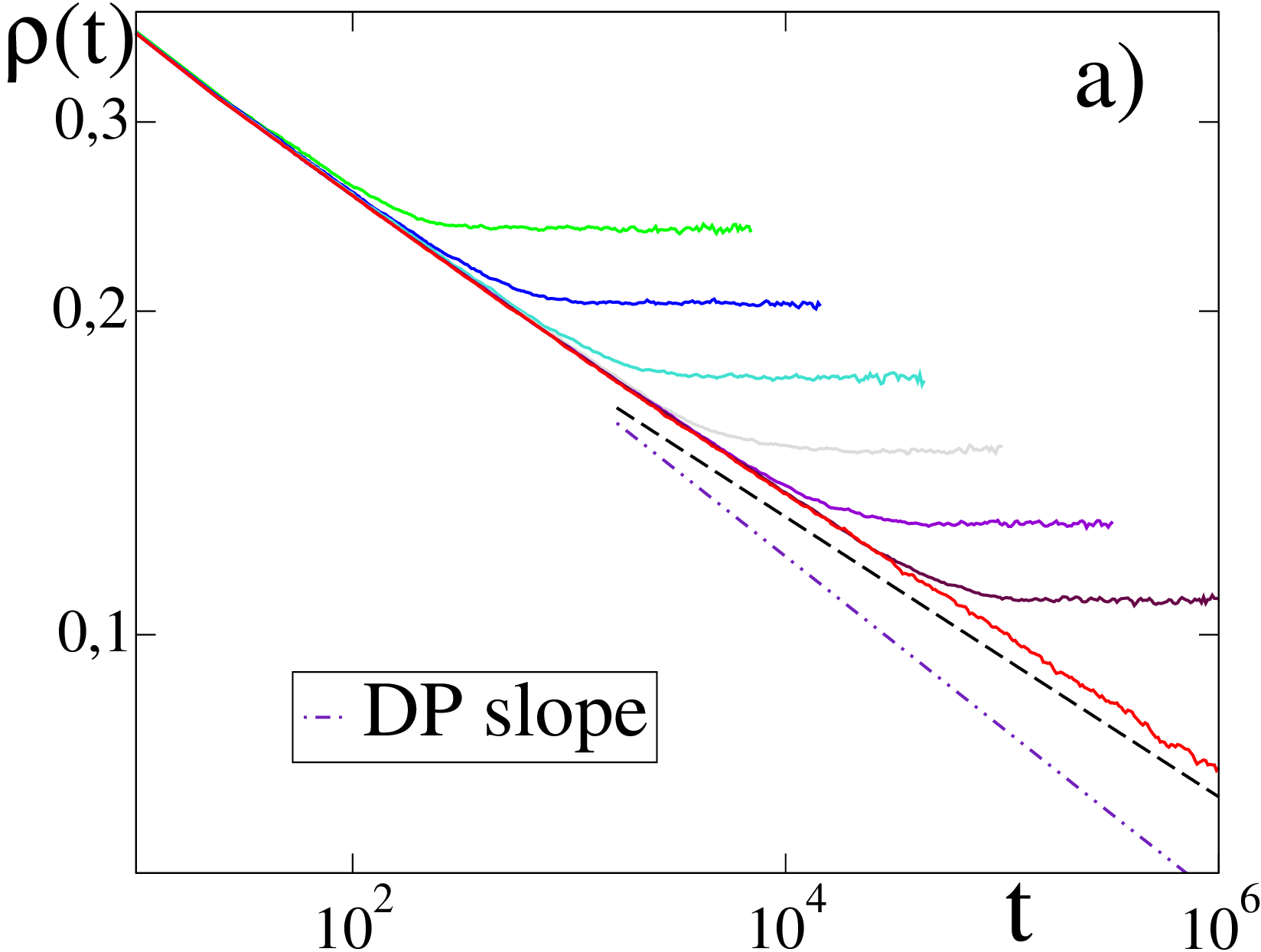}
\includegraphics[width=4.2cm]{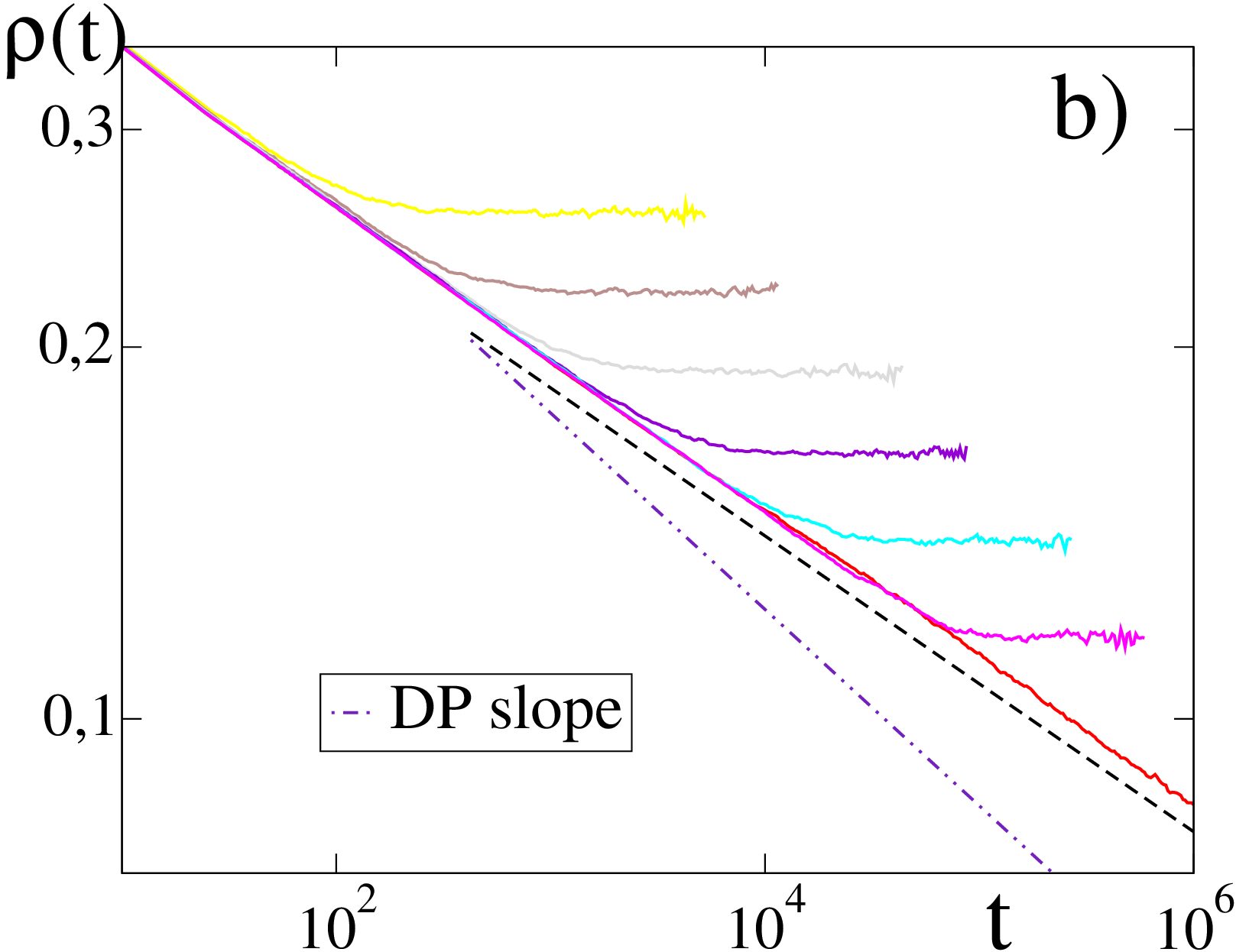}
\includegraphics[width=4.2cm]{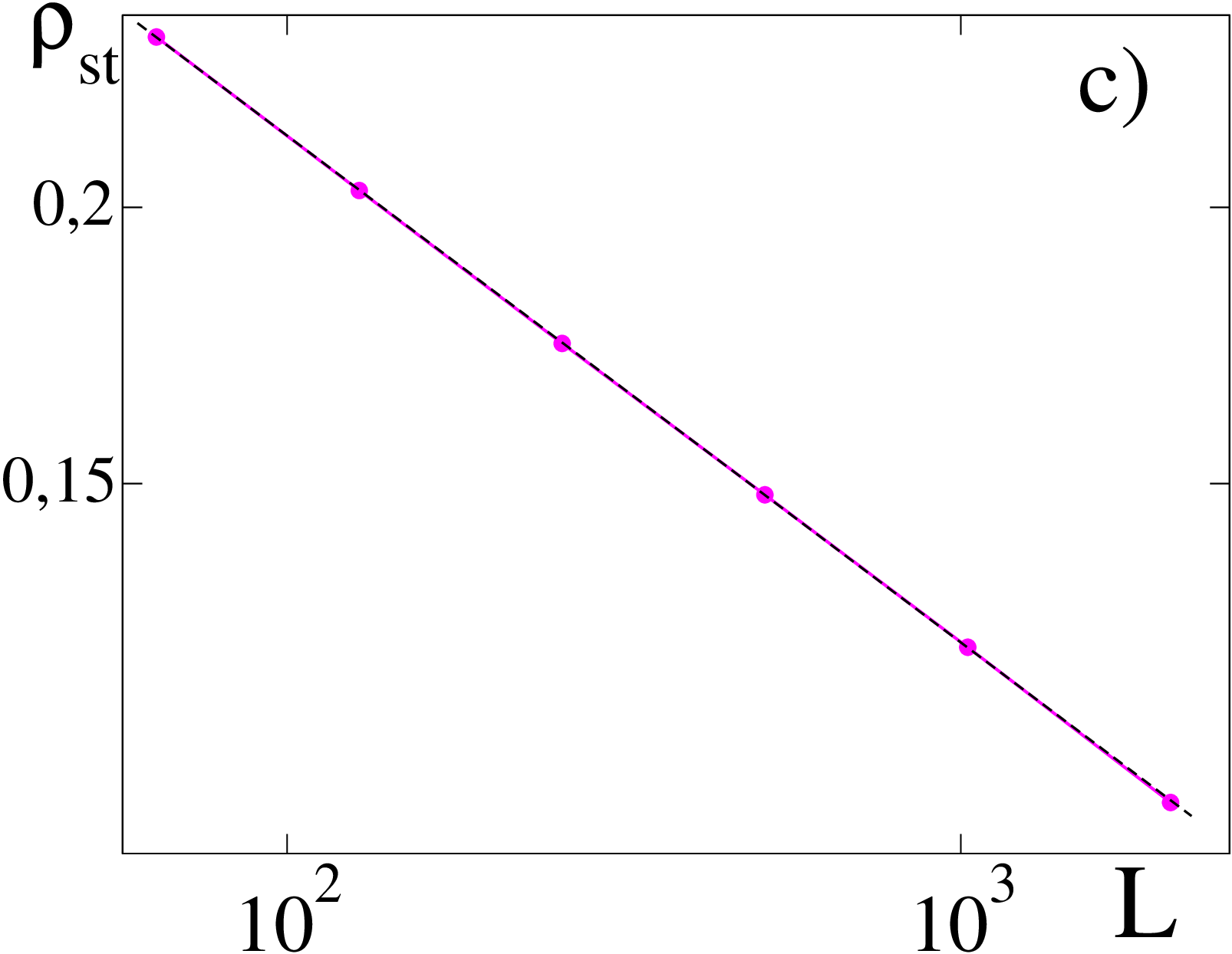}
\includegraphics[width=4.2cm]{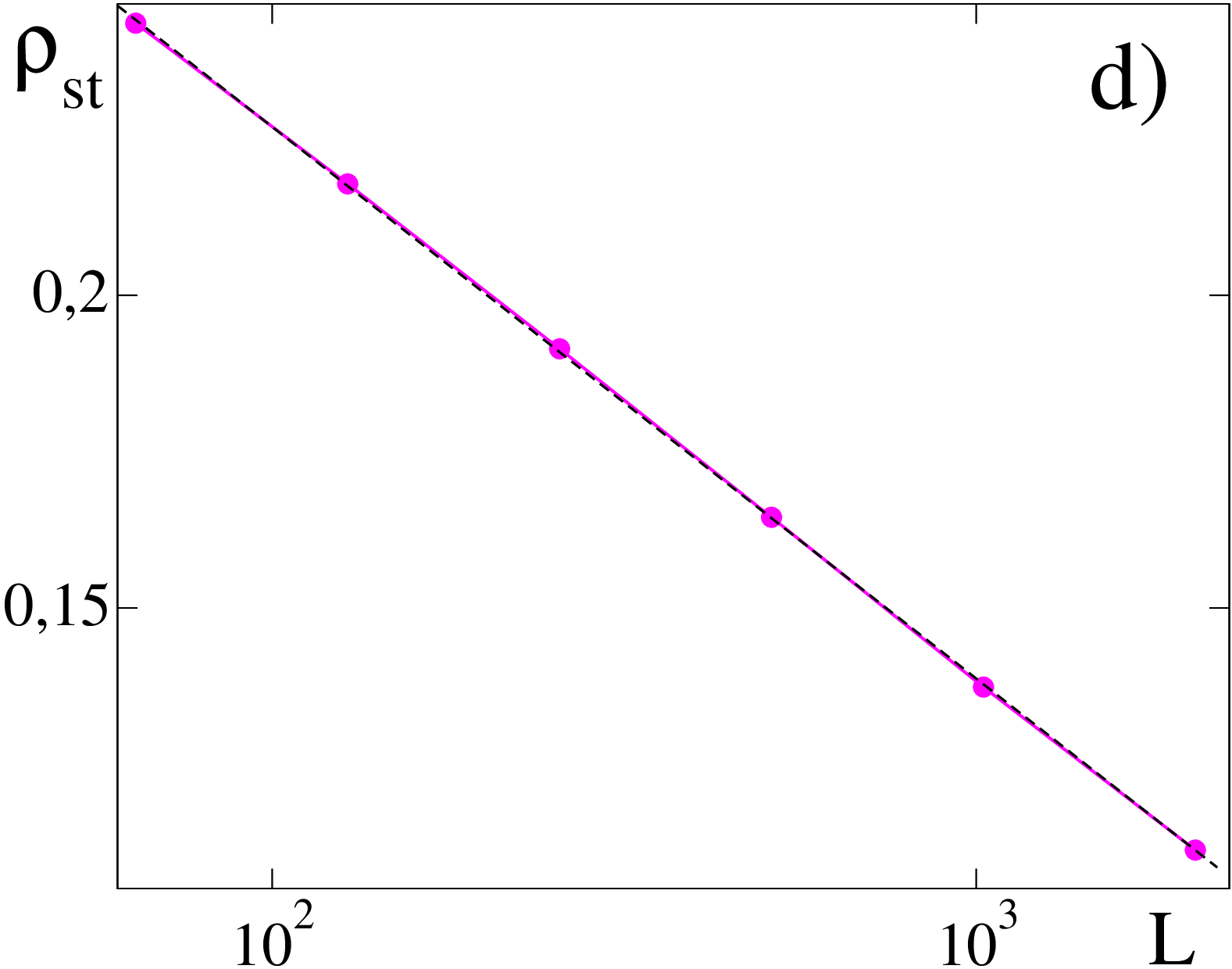}
\end{center}
\caption{(a-b) Log-log plot of the time-decay of the order parameter (activity
density) for different system sizes (from top to bottom: $L=64, 128,
256, 512, 1024, 2048$ and $L=2^{18}$) and for the (a) stochastic rule
and (b) the deterministic one, at their corresponding critical points
$p_{\rm c}=0.84937(2)$ and $p_c=0.76750(3)$. From the slopes, we
determine $\theta=0.120(8)$ and $\theta=0.115(8)$ respectively (DP
slopes are plotted for comparison). In (c-d) we plot the saturation
values at the previously determined critical points for the stochastic
(c) and deterministic (d) rules respectively. From the scaling at the
critical point we determine ${\beta\over\nu_{\perp}}=0.22(1)$ for both
of them.}
\label{f1}
\end{figure}

We have implemented two different versions in which each toppling
event redistributes two grains to the two nearest-neighbor sites,
either randomly (stochastic redistribution rule) or regularly, with
one grain onto each neighbor (deterministic rule).  The methodology
followed is standard for absorbing phase transitions: first, the
critical point is determined by studying the decay of activity from
some initial condition varying $p$ in a large system: for large $p$
values, activity saturates (active phase), while for small $p$
activity vanishes (absorbing phase). At the critical point, separating
these two phases, $p_{\rm c}$, activity decays asymptotically as a
power-law with the critical exponent $\theta=\beta/\nu_{\|}$.  For the
stochastic and the deterministic rules, we find, respectively, $p_{\rm
c}=0.84937(2)$ with $\theta=0.120(8)$ and $p_c=0.76750(3)$ with
$\theta=0.115(8)$ (Fig.~\ref{f1}a,b). These estimates of $\theta$ are
in good agreement with the best evaluations for the Manna class in one
dimension, {\it i.e.}  $\theta =0.125(2)$
\cite{DCM}, and clearly incompatible with the DP value $\theta 
\approx 0.159$.

Next, using the critical value determined above, the variation of the
stationary saturation value of activity at the critical point for
smaller system-sizes is recorded. From the expected scaling law
$\rho_{\rm st}(p=p_{\rm c})\sim L^{-{\beta\over\nu_{\perp}}}$, we
determine ${\beta\over\nu_{\perp}}=0.22(1)$ for both the stochastic
and the deterministic rule as shown in Fig.~\ref{f1}c,d. Again, this
value is in good agreement with available estimates for the C-DP/Manna
class ${\beta\over\nu_{\perp}}=0.215(5)$ \cite{KC,Ramasco}, and
incompatible with the DP value ${\beta\over\nu_{\perp}} \approx
0.252$.

We have also performed spreading experiments
\cite{Review_Haye,Review_Granada} (not shown) by following the  standard procedure: 
we perturb a natural absorbing state (one generated by the system
dynamics) to generate a small amount of localized activity and analyze
how it spreads out at the previously determined critical point. We
measured $\eta=0.39(3) $, $\delta=0.167(5) $, and $z=1.39(3) $
exponents for the number of active sites, surviving probability, and
average square-radius critical, respectively
\cite{Review_Haye,Review_Odor,Review_Granada}. These values are in good 
agreement with the best estimations for the Manna class \cite{Lubeck}
and differ from their corresponding DP values ( $\eta
\approx 0.313 $, $\delta \approx 0.159 $, and $z \approx 1.258 $
\cite{avalanches}).

\section{Numerical study of Langevin equations}

After revisiting microscopic sandpile models with sticky grains, their
critical behavior appears to be fully compatible with those of the
C-DP/Manna class. We now turn to a study of the coupled Langevin
equations proposed by Mohanty and Dhar to describe their
coarse-grained dynamics.  The stochastic equations proposed in
\cite{MD} are Eqs.~\ref{sde_Manna}, {\it i.e.} those of the C-DP/Manna
class, except for the coupling term $\omega \rho E$ which is replaced
by $\omega \rho \Theta(E-\rho-E_{\rm c})$ where $E_{\rm c}$ is the
(microscopic) toppling threshold.  The presence of this microscopic
feature and of the step function $\Theta$ is surprising in so far as
Langevin equations are usually understood as resulting from some
coarse-graining of microscopic dynamics. In particular, the step
function is unlikely to be a robust mesoscopic description, as it will
be modified (probably transformed into a smoother function) upon
coarse-graining.

Discontinuous functions are notoriously difficult in the framework of
renormalization group analysis. Moreover, even the ``simple''
equations (\ref{sde_Manna}) resist standard perturbative
renormalization attempts \cite{Fred}. The only available strategy then
is direct numerical integration. The presence of the (square-root)
multiplicative noise term makes this a priori difficult
\cite{Dickman_scheme}, but this technical difficulty was recently
circumvented by the fast and quasi-exact sampling method introduced in
\cite{DCM}.

We have used this scheme to numerically integrate the equations
proposed by Mohanty and Dhar. These simulations yielded two sets of
results: following the protocol recalled above, we studied the
absorbing phase transition observed when varying the linear
coefficient. We also introduced a local effective ``mass'' coefficient
\begin{equation}
a_{\rm eff} ({\bf x},t) = a + \omega ~
\Theta(E({\bf x},t)-\rho({\bf x},t)-E_{\rm c}),
\label{masa}
\end{equation}
and studied its behavior upon coarse-graining in numerical simulations
to analyze the relation between the two sets of Langevin equations.

\subsection{Phase transition}

\begin{figure}[t]
\begin{center}
\includegraphics[width=4.2cm]{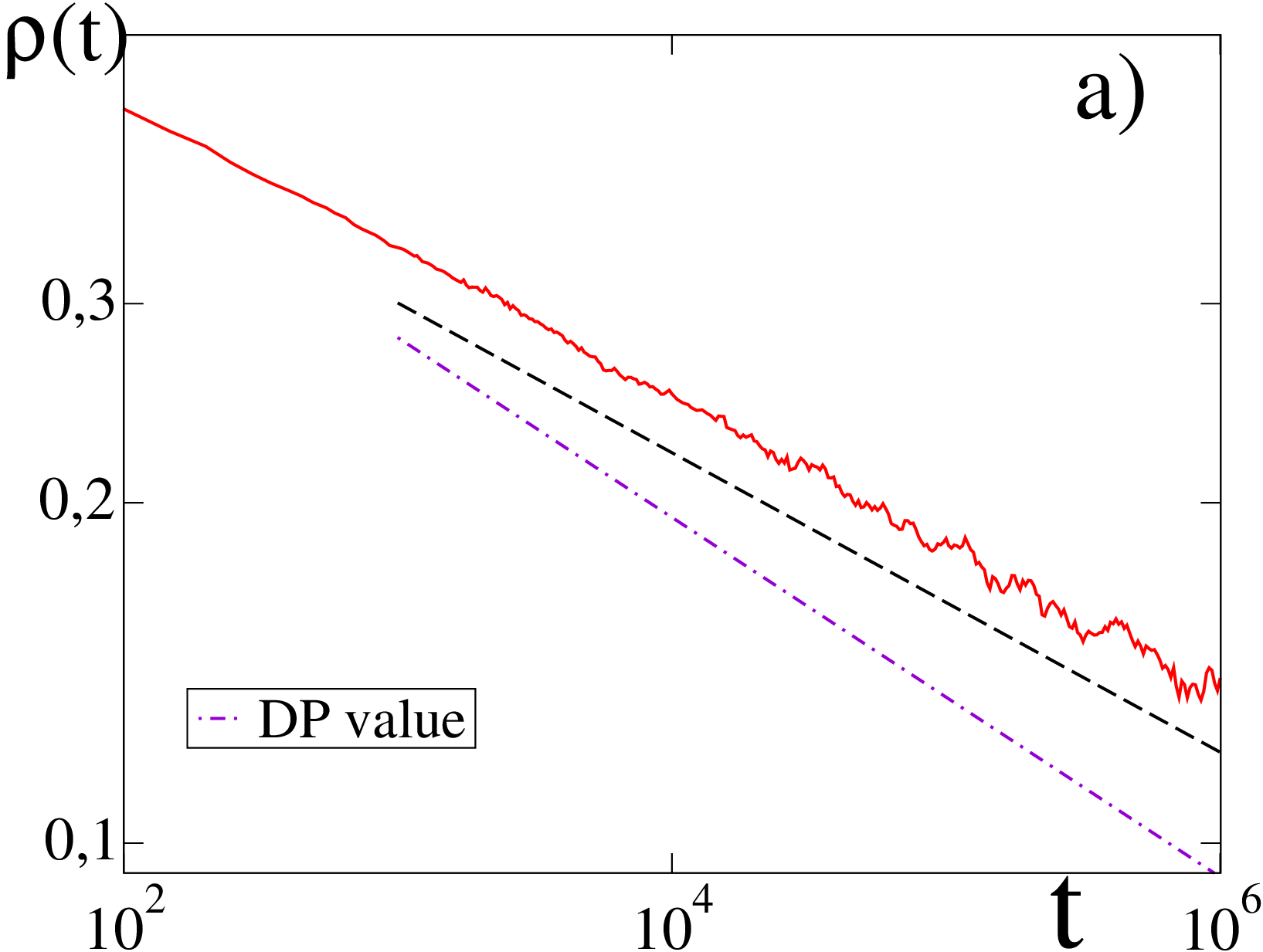}
\includegraphics[width=4.2cm]{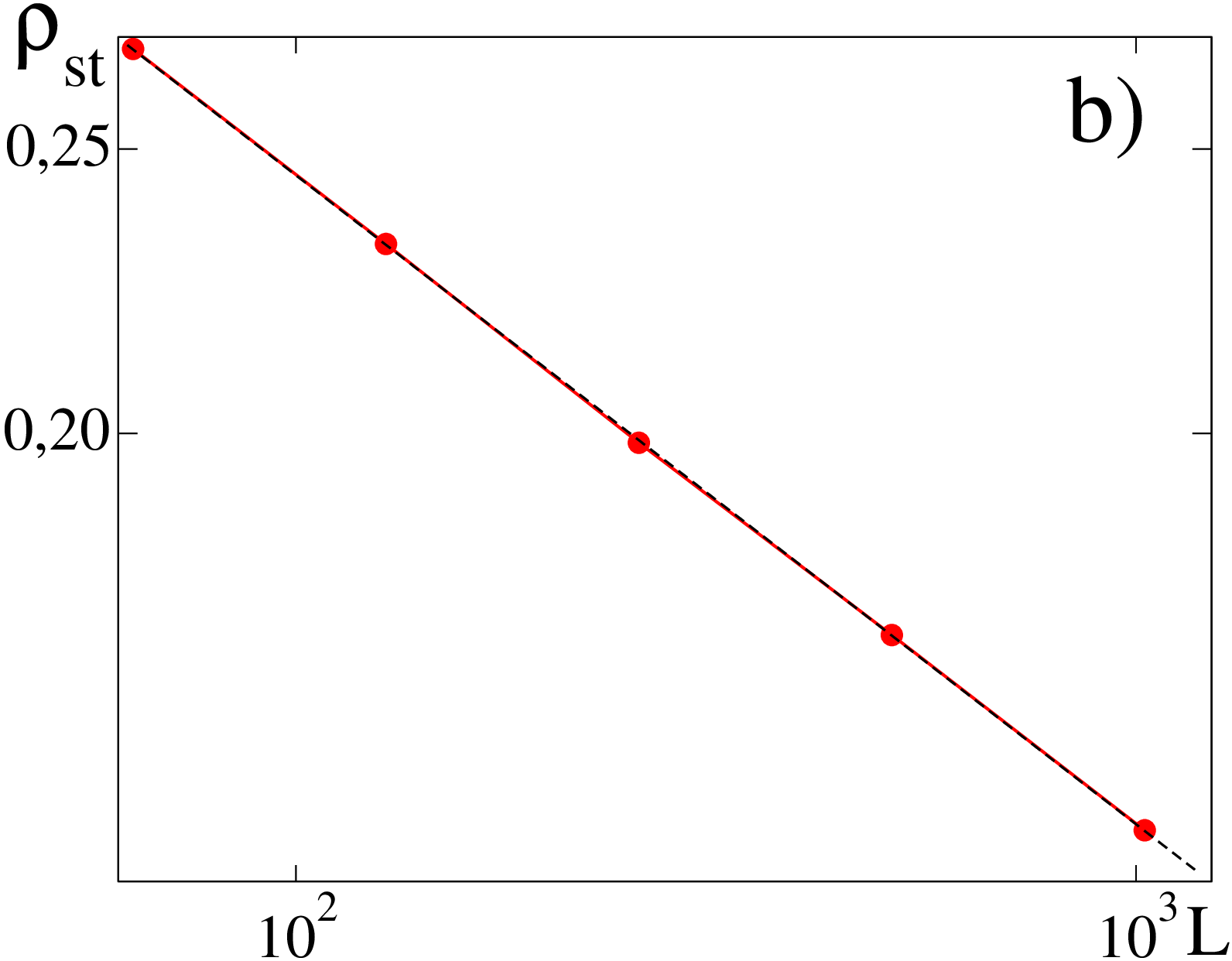}
\end{center}
\caption{Direct numerical integration of the Langevin equations 
proposed in \cite{MD}. (a) Decay experiments at the critical point
$a=0.72308$ (other parameter have been fixed as: system size
$L=2^{20}$, $\langle E(x,t=0) \rangle =0.5$, $b=1$, $h_c=0.5$,
$D=D_E=0.25$, $\omega^2=\sigma^2=2$, integrated with time-step
$dt=0.25$).  (b) Finite size scaling at criticality (more details in
the text).}
\label{f2}
\end{figure}

Starting from an homogeneous, active, initial condition, we studied
the time-decay of spatially-averaged activity varying the control
parameter $a$. As expected, algebraic decay is found at the critical
value separating exponential decay (absorbing phase) from saturation
(active phase).  The estimated decay exponent $\theta=0.130(5)$ is in
perfect agreement with the C-DP/Manna class value (Fig.~\ref{f2}a). At
the critical point, $a=0.72308$, the scaling of the stationary
activity for finite size systems yields the estimate
$\beta/\nu_\perp=0.22(1)$, (Fig.~\ref{f2}b) again in agreement with
the C-DP/Manna value. These estimates are thus incompatible with the
DP class values. Also, spreading experiments (not shown) fully confirm
this result.

\begin{figure}[t]
\vspace{0.45cm}
\begin{center}
\includegraphics[width=4.2cm]{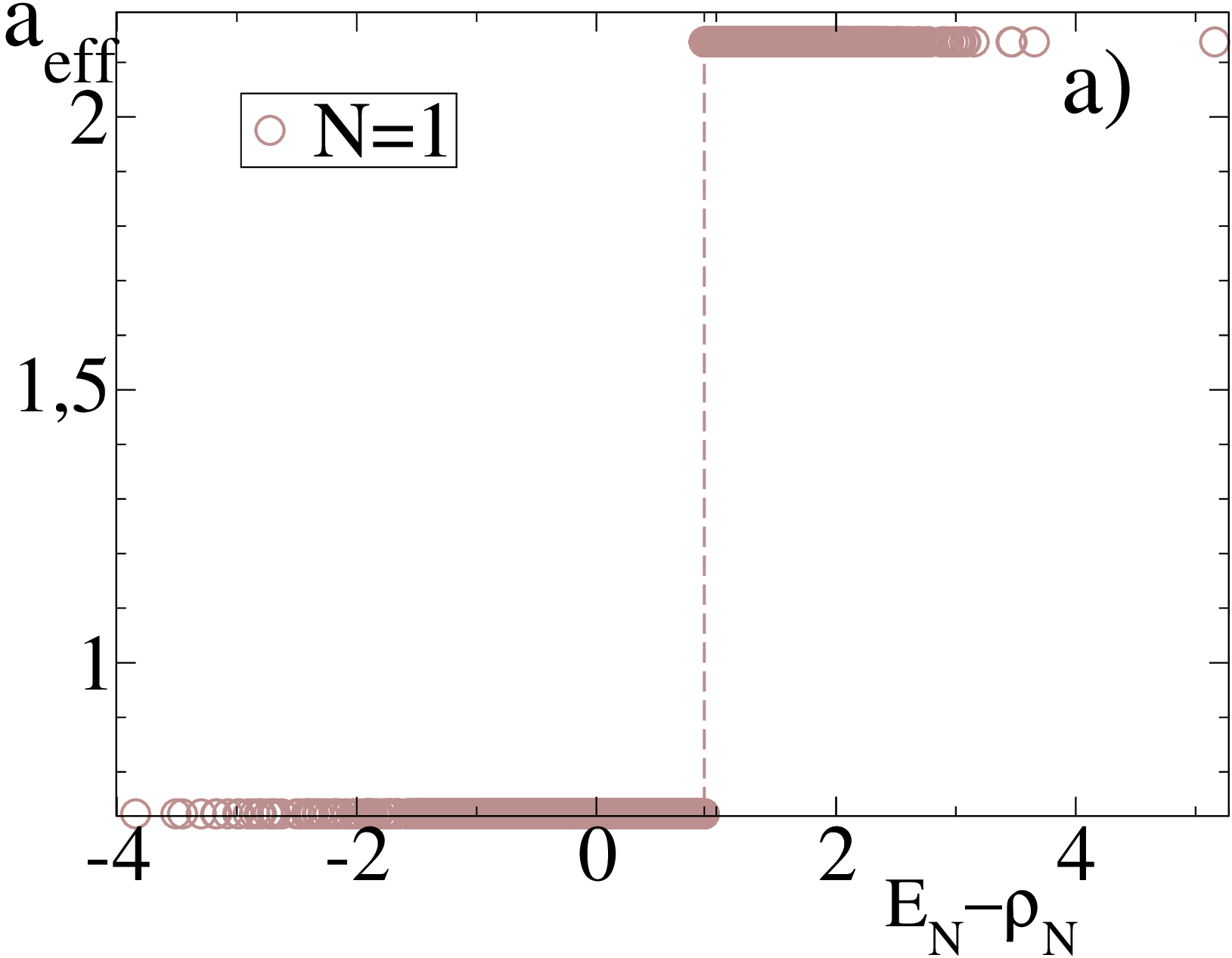}
\includegraphics[width=4.2cm]{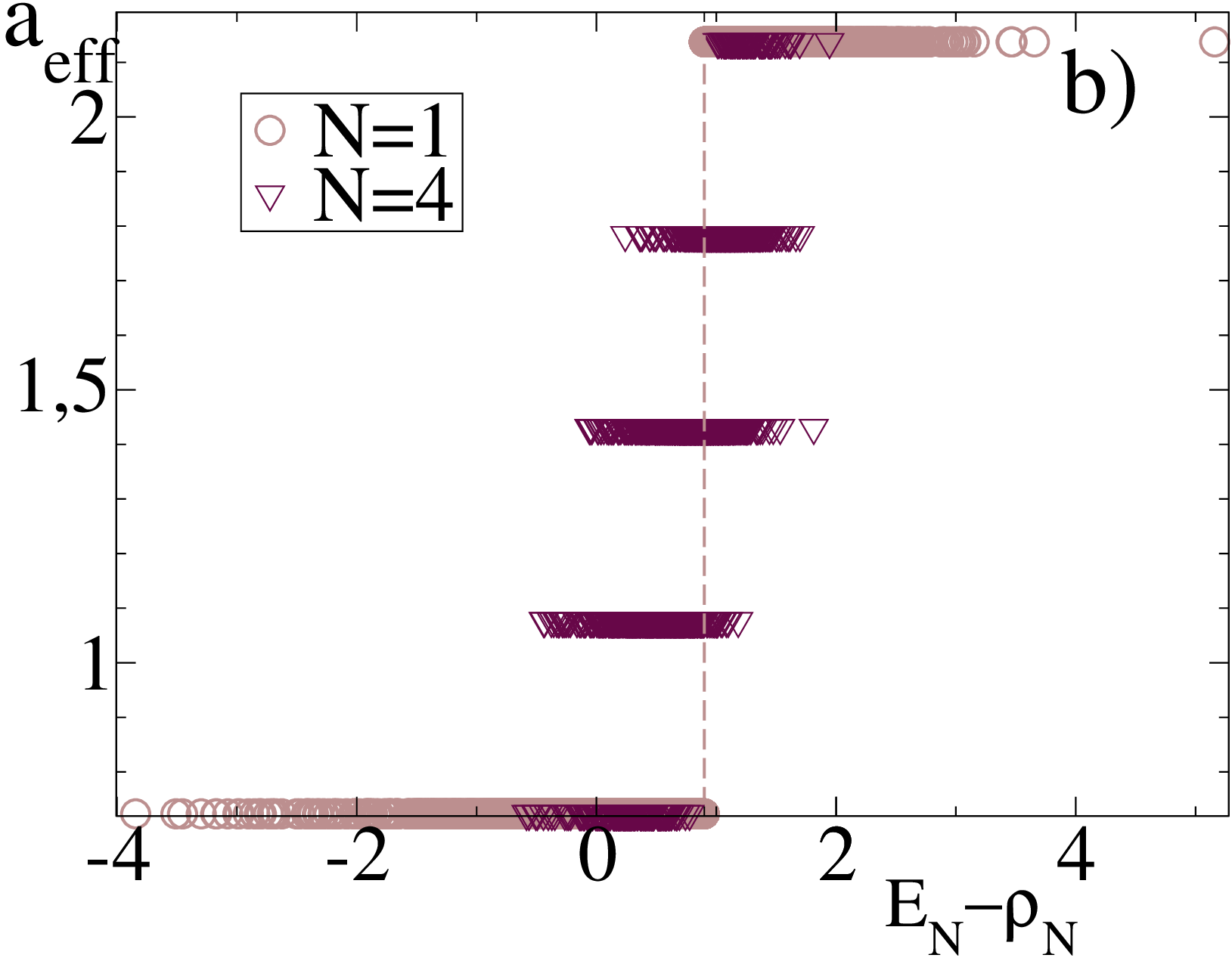}
\includegraphics[width=4.2cm]{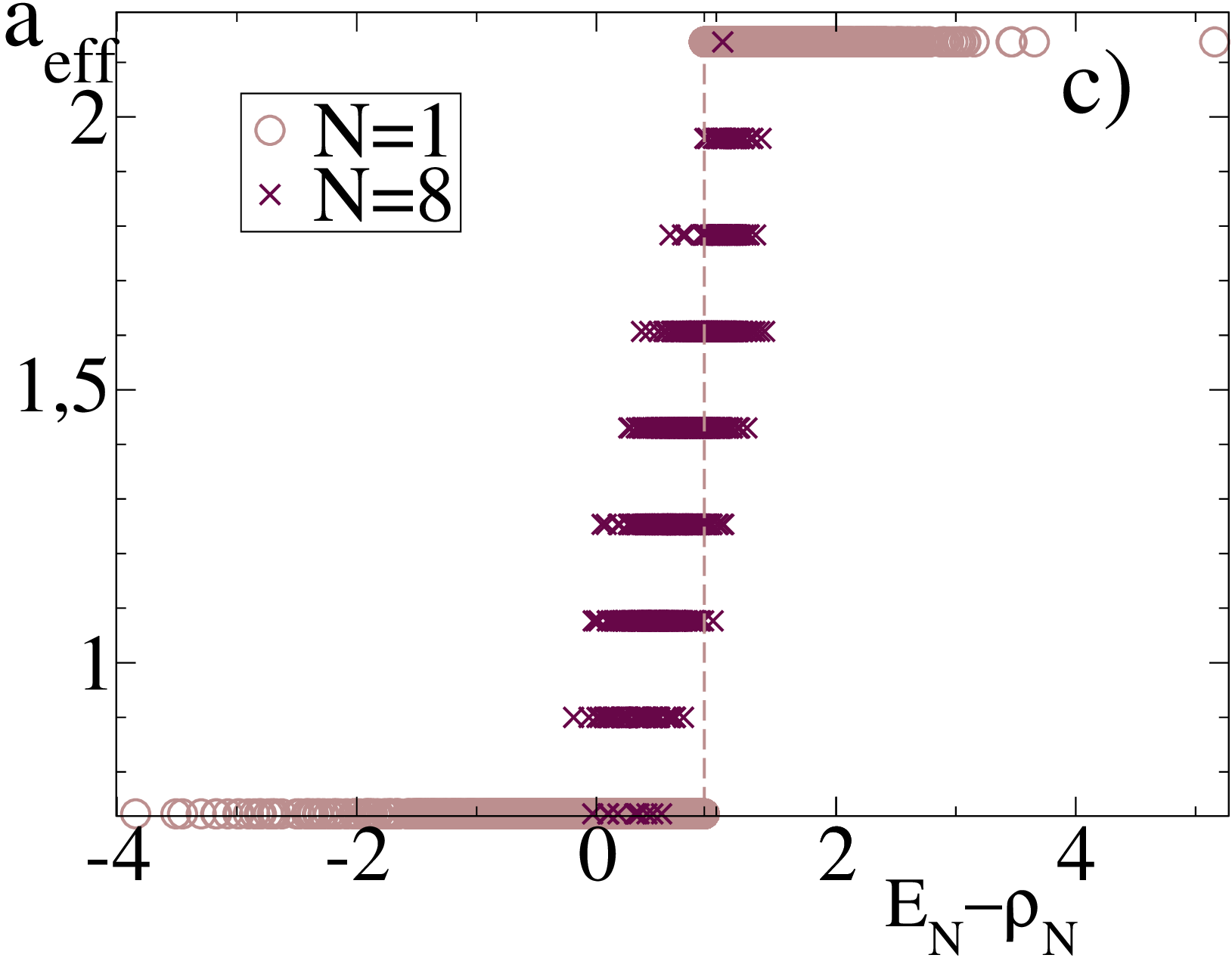}
\includegraphics[width=4.2cm]{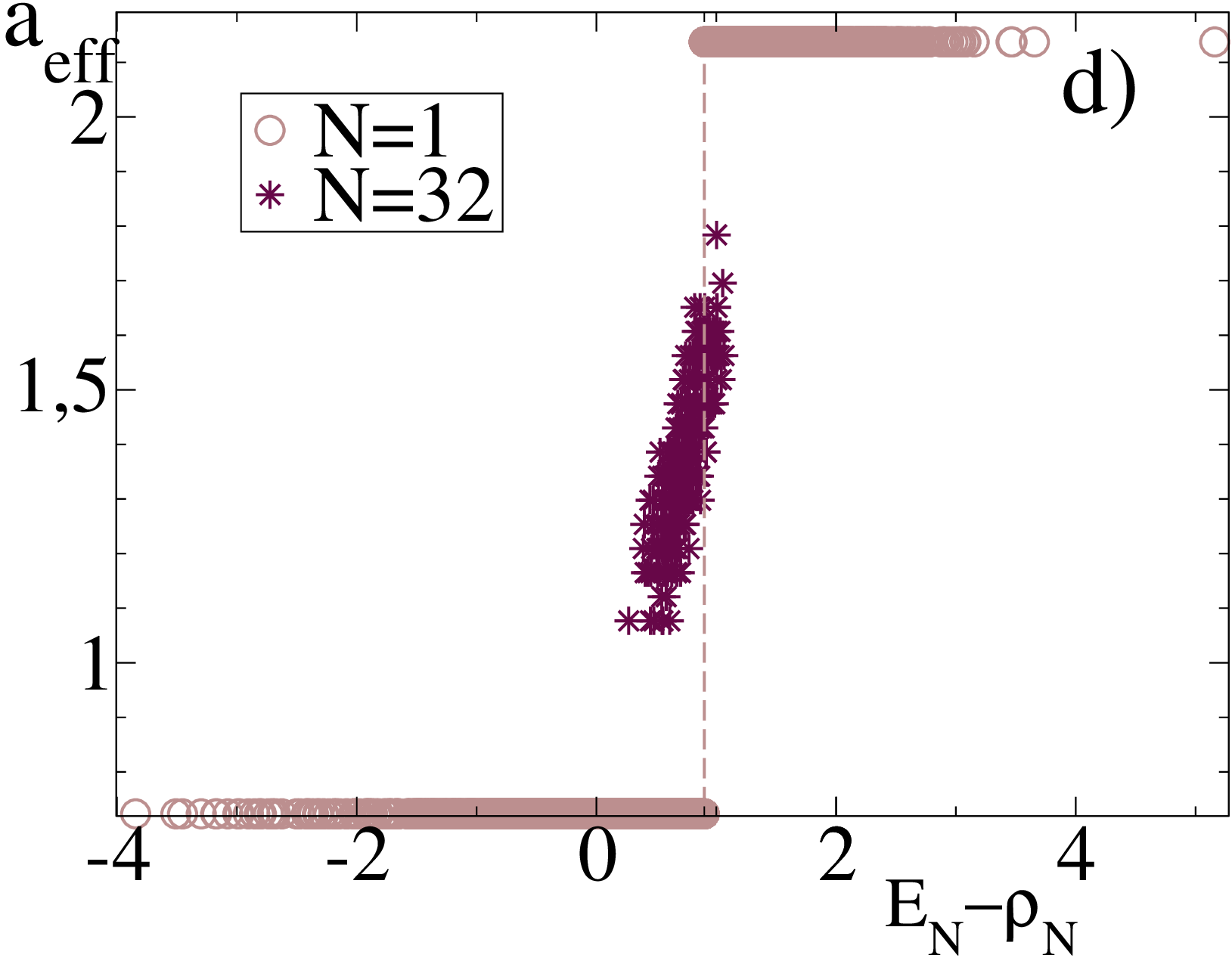}
\end{center}
\caption{\footnotesize{Effective mass as defined by Eq.~(\ref{masa})
as a function of the field difference averaged in Kadanoff blocks of
size $N$, for $a=0.72313$ in the active phase and $\omega=\sqrt{2}$.
The vertical line corresponds to the threshold value
$h_c=0.9$. Observe that the larger the block size, the smoother the
effective-mass dependence on the coarse-grained field difference.}}
\label{f3}
\end{figure}

\subsection{Numerical coarse-graining}

The above results are easily understood when observing the behavior of
the Mohanty-Dhar Langevin equations coarse-grained numerically.  To do
this, we build scatter plots of $\langle a_{\rm eff}\rangle_N$ vs the
field difference $\langle E-\rho\rangle_N$, where the averages are
taken on Kadanoff-blocks of length $N$.  For $N=1$ (Fig.~\ref{f3}a),
we obviously observe the $\Theta$-function form.  For $N=4$
(Fig.~\ref{f3}b), the effective coupling can take intermediate
discrete values between $a$ and $a+\omega$, depending on the number of
above-threshold microscopic sites in the block. When the block size is
larger and larger (Fig.~\ref{f3}c,d), a smooth function gradually
appears. By retaining just the two leading terms in a Taylor expansion
of such an analytical function around the origin, we recover, at a
coarse-grained level, the Langevin equation for the Manna class
Eqs.(\ref{sde_Manna}), i.e. a linear coupling term and a correction to
the linear term, $a$ in Eqs.(\ref{sde_Manna}).  Higher order terms in
the Taylor expansion can be argued to be irrelevant from standard
naive power counting arguments. Therefore, it is not surprising that
the set of Langevin equations including a Heaviside $\Theta$ function
should exhibit the same asymptotic behavior as the original C-DP/Manna
class Langevin equations, Eqs.(\ref{sde_Manna}).

\section{Conclusion and discussion}

To sum up, introducing ``stickiness'' in isotropic sandpile models
does not change their universality class, which remains that of the
Manna model. We reached this conclusion via numerical simulations of
microscopic models and Langevin equations proposed in \cite{MD}. We
showed in addition that these Langevin equations ``flow'' towards
those of the Manna class under some numerical coarse-graining
procedure.

One can wonder what is the reason why this conclusion does not hold
for the {\it directed} (or anisotropic) sandpiles studied also by
Mohanty and Dhar in \cite{MD} (see also \cite{Tadic}), which they
proved to be in the DP class. These directed models, defined on a
two-dimensional lattice include an isotropic direction and a fully
anisotropic one, in the sense that sand goes ``downwards'' in that
direction but not ``upwards''.  This makes it possible to map the
problem on DP in $(1+1)$ dimensions, {\it i.e.} the anisotropic
dimension can be taken as ``time''.  The local conservation of energy
is present also in these models, but ``local'' here means in
``space-time'' neighborhoods, while {\it energy is not conserved} in
the isotropic spatial direction. Hence, Manna behavior does not
appear, while DP scaling, the usual one in the absence of spatial
energy conservation, is expected to emerge, as indeed was proved in
\cite{MD}

M.A.M. acknowledges financial support from the Spanish MEC-FEDER,
project FIS2005-00791, J.A.B. acknowledges support under a fellowship
from the MEC. We also acknowledges Deepak Dhar for useful comments and
the initial suggestion that led to this study.

\end{document}